# Ultra-high-speed coherent anti-Stokes Raman spectroscopy with a hybrid dual-comb source

Tianjian Lv[1], Bing Han[1], Ming Yan[1,2,*], Zhaoyang Wen[1], Kun Huang[1,2], Kangwen Yang[3], and Heping Zeng[1,2,4,5,*]

1. State Key Laboratory of Precision Spectroscopy, East China Normal University, Shanghai 200062, China
2. Chongqing Key Laboratory of Precision Optics, Chongqing Institute of East China Normal University, Chongqing 401120, China
3. Shanghai Key Lab of Modern Optical System, University of Shanghai for Science and Technology, Shanghai 200093, China
4. Jinan Institute of Quantum Technology, Jinan, Shandong 250101, China
5. Chongqing Institute for Brain and Intelligence, Guangyang Bay Laboratory, Chongqing 400064, China.

\* Corresponding authors, myan@lps.ecnu.edu.cn, hpzeng@phy.ecnu.edu.cn

**Abstract**

Coherent anti-Stokes Raman scattering (CARS) spectroscopy with time-delayed ultrashort pulses and a single-pixel photodetector has shown great potential for spectroscopic imaging and transient studies in chemistry and biological research. However, those systems rely on mechanical delay lines or two asynchronous optical combs with inflexible repetition frequencies, technically limiting their acquisition speeds. Here, we demonstrate a hybrid dual-comb CARS system involving a broadband fiber laser and a highly-flexible, frequency-modulated electro-optic comb. We achieve multiplex CARS spectra (2800-3200 cm$^{-1}$), with a moderate resolution (30 cm$^{-1}$), at a maximum refresh rate of 1 MHz, limited by the radio-frequency synthesizer we use. Fast spectroscopic CARS imaging is demonstrated for liquid mixtures. Our system enables spectral measurements in the high-wavenumber C-H stretching region at a record speed that is an order of magnitude higher than state-of-the-art systems, which may open up new opportunities for fast chemical sensing and imaging.

**TOC Graphic**

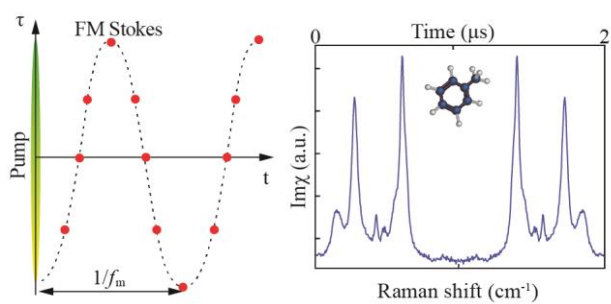

# 1. Introduction

Studies on highly dynamic systems and fast, non-repetitive events, as encountered in, e.g., transient chemical reactions,[1] high-throughput flow cytometry,[2] and *in vivo* biological imaging,[3] to name a few, are prevalent in analytical chemistry and biomedical research, and constantly call for novel high-speed spectroscopic approaches. Broadband CARS spectroscopy is a powerful tool that captures a wealth of molecular vibrational information for discriminating spectrally overlapped chemical species and subtly different pathological states of biological systems.[4-10] Also, benefiting from coherently enhanced Raman signals, it allows for fast detection. As such, broadband CARS holds great promise for integrating inherent chemical specificity and high-speed measuring capability on a single platform, which is crucial for the studies mentioned above. So far, broadband or multiplex CARS spectra have been routinely obtained with camera-based dispersive spectrometers. But these systems suffer from a camera's ms-level read-out time,[4] restricting their applicability in many of the above scenarios. Other methods, such as hyperspectral CARS[11] and single-beam CARS,[12,13] are limited by laser tuning speeds or pulse shaping rates. Therefore, obtaining broadband CARS spectra in an ultra-rapid way remains challenging.

Recent Fourier-transform (FT) spectroscopy advances lead to new opportunities for high-speed CARS measurements.[14-18] In FT CARS, ultrashort pulses with varying time delays are harnessed to stimulate multiple Raman vibrations simultaneously, yielding interferometric anti-Stokes signals that are detected, in the time domain, by a fast single-pixel photodetector. The rapid spectral acquisition has been achieved at dozens of kHz by scanning the pulse delays with either a resonant scanning mirror[15] or a polygonal scanner.[16] But, the mechanical delay lines have inertia that ultimately limits their speeds and need post-phase corrections with extra efforts in data computing.

Alternatively, FT CARS with two frequency combs of slightly different repetition frequencies, i.e., dual-comb CARS,[19-30] has been developed to overcome these limitations. This scheme, without moving parts, reduces single-shot measurement times to a few µs or less.[19] In its original version, it is limited by low refresh rates (varying from Hz to kHz levels) and small duty cycles (~$10^{-3}$) due to the mismatch between comb repetition periods (~ns) and Raman coherence times (~ps). Increasing comb repetition frequencies offers a straightforward solution,[20] but it causes other unwanted problems, such as reduced signal-to-noise ratio (SNR), high average optical powers, and photodamage. An elegant approach to address the issues is recently demonstrated by a quasi-dual-comb scheme, where the cavity length of one of the combs is rapidly modulated by changing its pump current.[24] Consequently, the dual-comb refresh rate is improved up to 100 kHz with a duty cycle close to 100%, but it sacrifices the spectral resolution (117 cm$^{-1}$) due to the tradeoff in FT spectroscopy. Also, the system relies on two high-repetition-rate combs at ~1 GHz, with the disadvantage of pulse energy required for nonlinear measurements. Similar dual-comb schemes have also been demonstrated for spectral-focusing (SF) CARS,[25-28] a time-domain hyperspectral technique. With two 100-MHz combs, one of which has a repetition frequency being modulated by an intra-cavity electro-optic (EO) modulator, coherent Raman spectra (>200 cm$^{-1}$ bandwidth) at a high resolution (10 cm$^{-1}$) are obtained at a refresh rate up to 40 kHz, enabling high-throughput flow cytometry.[28] However, a limiting factor of its speed would be the response frequency of the intracavity EO modulator.[28] More essentially, these dual-comb systems rely on mode-locked lasers with optical cavities that are environmentally sensitive and have little flexibility in their repetition frequencies. EO comb generation offers a simple and robust source with high flexibility and tunability. Dual-comb CARS with two EO combs has been reported and achieves acquisition rates in a range of 10 to 50 kHz, which, however, operates at high repetition frequencies (~10 GHz).[29] So far, the excellent flexibility of EO combs has not been exploited for high-speed CARS measurements.

In this paper, we demonstrate a hybrid dual-comb CARS technique that combines the advantages of a highly flexible frequency modulation (FM) EO comb and a broadband fiber laser for ultra-high-speed CARS spectroscopy. More specifically, we directly convert a fast FM electric signal into a train of sub-ps optical pulses, together with an asynchronous temporally chirped fiber laser, enabling multiplex measurements (2800-3200 cm$^{-1}$ with a 30 cm$^{-1}$ resolution) at a speed up to 1 MHz, which is an order of magnitude faster than the state-of-the-art systems.[24, 28] Therefore, our technique may lead to new opportunities for chemical sensing and imaging applications.

**2. Basic concepts**

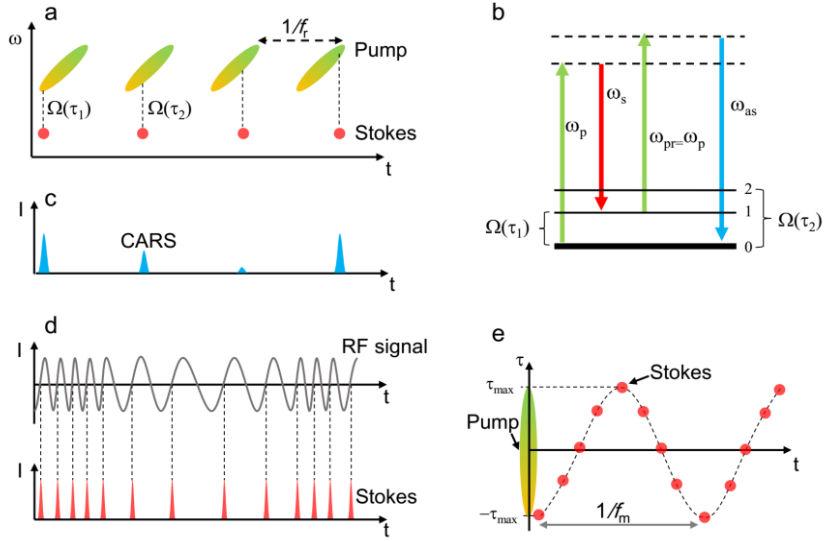

**Figure 1.** Diagrams of frequency modulated dual-comb CARS. (a) The temporal-spectral diagram of pump and Stokes pulses. (b) The energy diagrams of CARS. (c) The CARS signals generated in the time domain. (d) A FM signal is transformed into electrical pulses at the zero-crossing points of the rising edges. (e) The relationship between the relative pump-Stokes time delay, $\tau$, and the measurement time, t.

Our method is based on time-encoded asynchronous optical sampling (AOS). As shown in Figure 1a, a broadband ultrafast laser (center carrier angular frequency: $\omega_P$), which is temporally stretched with a linear chirp rate of $\beta$, is used as a pump (probe) beam, and a narrowband pulse laser as a Stokes beam at $\omega_S$. In a typical AOS process, the two lasers have slightly different repetition rates of $f_r+\Delta f$ and $f_r$, respectively. A train of paired pulses (each pair contains a pump and a Stokes pulse) is thereby obtained. The relative delay ($\tau = N \cdot \Delta f / f_r^2$) between the pulses in a pair increases linearly with the sequence number (N). Consequently, the relative delay is automatically scanned without the limitations of using mechanical scanners. Also, the instantaneous frequency difference (IFD) of the two lasers is a function of their relative time delay $\tau$, as $\Omega(\tau)=\omega_P-\omega_S+2\beta\cdot\tau$, provided the Stokes bandwidth is relatively small. When IFD at a certain delay ($\tau_1$) coincides with a Raman active vibronic mode (frequency: $\nu_{vib}$) of the molecules (i.e., $\Omega(\tau_1)=\nu_{vib}$), a coherent anti-Stokes (or CARS) signal, at $\omega_{as}=\omega_P+\Omega(\tau)=2\omega_P-\omega_S+2\beta\cdot\tau_1$, will be generated via a resonant four-wave mixing (FWM) process (Figure 1b). This blue-shifted light is detected with a fast photodetector.

The detector signal varies with the relative delay time, presenting a CARS spectrum encoded in the time domain (Figure 1c). The Raman span interrogated is determined by the pump spectral width and the resolution by a convolution between the effective bandwidth of the chirped pump laser and the Stokes bandwidth. The AOS process in Figure 1a is the same as that in the original dual-comb schemes.[19-23] The relative delay will continuously increase until a pump and a Stokes pulse coincide again (i.e., t=1/$f_r$). This causes drawbacks, including poor duty cycles and long dwell times (therefore low data refresh rates).

To overcome these drawbacks, we employ an electrical pulse generator to convert a sine-wave FM signal into a sequence of electrical pulses, which are then transformed into ultrashort optical pulses via an EO modulator for Stokes pulse generation (Figure 1d). The FM signal can be described as:

$$S(t) = A\cos[2\pi f_r t + \frac{\Delta f}{f_m}\sin(2\pi f_m t)] \qquad (1)$$

where $A$ is the signal amplitude, $f_r$ the carrier frequency, $f_m$ the modulation frequency, and $\Delta f$ is the frequency deviation. The FM signal (also the Stokes' repetition rate) oscillates between $f_r+\Delta f$ and $f_r-\Delta f$. Consequently, the relative time delay ($\tau$) between the Stokes and the pump (fixed at $f_r$) pulses varies with measurement time ($t$) as

$$\tau = -\frac{\Delta f}{2\pi f_m} \cdot \frac{1}{f_r} \cdot \cos(2\pi f_m \cdot t) + C \ . \qquad (2)$$

where $C$ is a constant from the integral of the sinusoidal FM function. As shown in Figure 1e, the pump-Stokes pulse delay is controlled in a way that the Stokes pulse oscillates around the pump pulse at the frequency $f_m$. Since in each modulation cycle (1/$f_m$) the Stokes pulse passes the pump pulse twice, the spectral acquisition speed is twice the modulation frequency (i.e., 2$f_m$). The relationship in the above equation is also experimentally verified (see Supporting Information). Radio-frequency (RF) synthesizers are commercially available for producing FM signals with $f_m$ at MHz level (or even higher), hence allowing for measurements at speed beyond the reach of the mode-locked dual-comb CARS systems. Also, with apt parameters (i.e., $f_r$, $f_m$ and $\Delta f$), the maximum delay, $\tau_{\max} = \frac{\Delta f}{\pi f_m} \cdot \frac{1}{f_r}$, can be equal to the pump pulse width so that 100% duty cycle can be achieved.

Although our method is technically equivalent to the quasi-dual-comb[24] or similar FM dual-comb schemes,[27, 28] it has several advantages due to the replacement of one of the mode-locked combs with an EO comb. Firstly, both $f_r$ and $f_m$ of an EO comb can be set arbitrarily (by digital synthesizers); while for a mode-locked comb, $f_r$ is strictly limited by its cavity length, and $f_m$ by the ways of the modulation applied (such as using a piezoelectric transducer or an EO modulator or pump current) and by the response time of the comb to the modulation. Secondly, the EO comb inherits the stability from the RF signal despite the comb's repetition rate being rapidly modulated, which is challenging to implement with mode-locked combs.[24, 27, 28] Also, compared to a mode-locked comb, the EO comb, without an optical cavity, has better disturbance immunity and robustness, which is preferred in practical use.

## 3. Methods

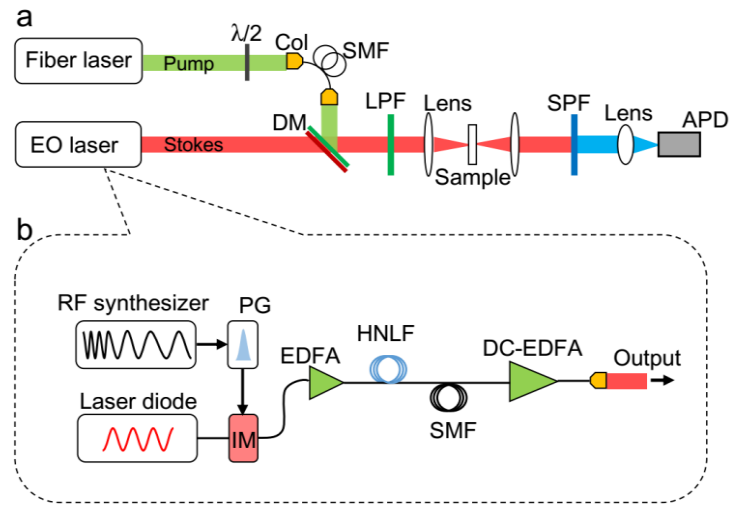

**Figure 2.** Experimental setup. (a) Hybrid dual-comb CARS setup. (b) Scheme for EO comb generation. Col, collimator; SMF, single-mode fiber; IM, intensity modulator; EDFA, erbium-doped fiber amplifier; HNLF, highly-nonlinear fiber; DCF, double-cladding fiber; DM, dichroic mirror; LPF, long-pass fiber; SPF, short-pass filter; APD, avalanche photodiode.

The experimental setup is depicted in Figure 2a. A user-customized mode-locked ytterbium-doped fiber laser system is employed for pump pulse generation. The system consists of a compact all-polarization-maintaining (PM) fiber oscillator that delivers self-starting, stable pulses at 1030 nm, with a stabilized repetition rate at $f_r^{(P)}$=60.5 MHz. Following the oscillator, a ytterbium-doped fiber amplifier (YDFA) boosts the average pulse power from 8 mW to 1 W. The YDFA output is temporally compressed to ~190 fs (full width at half maximum or FWHM) and then spectrally broadened in a ~7 cm photonic crystal fiber (PCF) before being sent out. The details regarding the fiber laser system are provided in Supporting Information. The laser output is coupled into ~4 m single-mode fiber (SMF) via an integrated fiber collimator to generate linearly chirped pump pulses.

Unlike prior dual-comb CARS systems[25-28] resorting to mode-locked lasers for Stokes pulse generation, our system utilizes a cavity-free EO comb to generate sub-ps Stokes pulses, featuring a highly flexible repetition rate. In Figure 2b, the EO comb starts from a seed laser, i.e., a laser diode (CBDX1-1-C, ID Photonics) emitting at 1550 nm. About 100 mW of the seed laser is fed to a fiber-coupled Mach-Zehnder intensity modulator (MXAN-LN-40, Photline; 30 GHz bandwidth) with a low half-wave voltage of ~6.4 V. To drive this modulator, an RF synthesizer (DSG3060, Rigol) generates a sinusoidal waveform, which is transformed, by an electrical pulse generator (LPP-100, LaserGen), into a sequence of low-time-jitter (<0.4 ps) 30 ps electric pulses, subsequently amplified by an RF power amplifier (40 GHz bandwidth). This electrical amplifier is integrated with the modulator. Consequently, a train of ~30 ps optical pulses is obtained at the intensity modulator output. The EO repetition rate ($f_r^{(S)}$), set by the RF synthesizer, is 302.5 MHz, giving rise to a pulse duty ratio of ~$10^{-3}$ and an average output power of ~100 μW. For CARS measurements, the RF synthesizer operates under FM mode.

Here, we set $f_r^{(S)}$ to the fifth harmonics of $f_r^{(P)}$, i.e., $f_r^{(S)}$=5·$f_r^{(P)}$=302.5 MHz, due to the following considerations. First, $f_r^{(S)}$ is within a comfortable regime (300-500 MHz) of the electrical pulse generator, regarding the pulse intensity stability and the pulse contrast ratio (>35 dB). Second, this does not affect the AOS process under FM mode, as long as temporal aliasing is avoided, which is satisfied as the designed maximum delay time ($\tau_{max}$ ~16 ps discussed in the results part) is less than the Stokes pulse repetition period ($1/f_r^{(S)}$ ~3.3 ns). In exchange, 80% of the Stokes pulses are discarded. One may solve this problem by reducing $f_r^{(S)}$ (e.g., using a pulse picker before optical power amplification) or increasing the pump's repetition rate ($f_r^{(P)}$) to match $f_r^{(S)}$.

A challenge in our scheme is to compress the 30-ps EO pulses into a sub-ps regime with enough pulse energy and intensity for CARS spectroscopy. To this end, the EO pulses are first amplified to 300 mW (single pulse energy ~1 nJ) by a lab-built erbium-doped fiber amplifier (EDFA) and then spectrally broadened in ~200 m normal-dispersion highly nonlinear fiber (HNLF). The HNLF has a zero-dispersion wavelength at 1550 nm and a linear dispersion slope of 0.03 ps/(nm$^2$·km) with a loss coefficient of 1.5 dB/km and a nonlinear coefficient of ~10W$^{-1}$km$^{-1}$. For dispersion management, the HNLF output is sent to ~150 m single-mode fiber (SMF), followed by a lab-built double-cladding fiber EDFA (DCF-EDFA) for compensating the pulse energy loss. The DC-EDFA output, with sub-ps pulse duration and ~0.5 W of maximum average power, is launched into free space through a fiber collimator and ultimately used as the Stokes beam.

For CARS measurements, the pump beam passes through a half-wave plate and is then spatially overlapped with the Stokes beam on a dichroic mirror (DMLP1180, Thorlabs). After that, a long-pass filter (FELH1050, Thorlabs; cutoff at 1050 nm) is used for spectral filtering. The filter is slightly tilted to shift its cutoff wavelength to 1035 nm for extending the Raman shift to a higher wavenumber (3200 cm$^{-1}$) and blocking the short-wavelength components of the pump light (which may overlap with CARS signals). The combined beams are then focused onto a liquid sample (in a 1-mm-thick cuvette) with a lens of 4.5 mm focal length (C230TMD-C, Thorlabs). The total average power on the sample is 280 mW (~80 mW for the pump and ~200 mW for the Stokes). The anti-Stokes radiation is forward-collected and collimated with another lens (C230TMD-B, Thorlabs) and then focused onto a 1 GHz-bandwidth avalanche photodiode (APD; C5658, Hamamatsu) after a short-pass filter (FESH0950, Thorlabs; cutoff at 950 nm). The APD output signal is digitized by a 12 bits oscilloscope (Teledyne Lecroy HDO6104A, 1 GHz bandwidth) for data acquisition. Note that the fiber laser, the EO comb, and the oscilloscope are disciplined to a hydrogen maser (stability: 10$^{-13}$ at 1s).

## 4. Results and discussions

*4.1 Characterizations of laser sources*

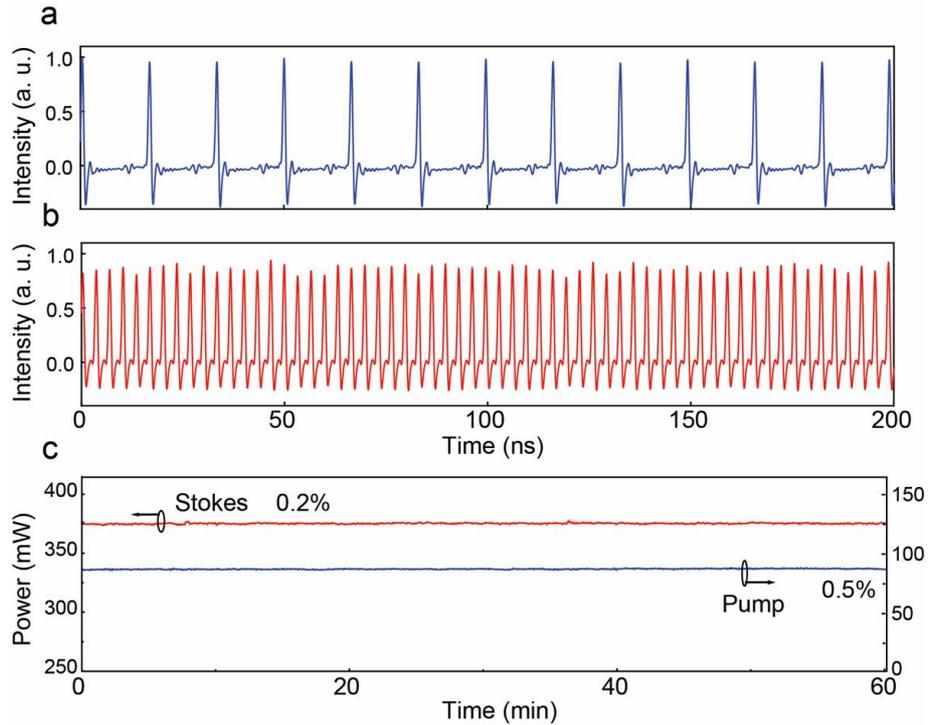

**Figure 3.** Characterization of the pump and Stokes lasers. (a) The pump pulse train and (b) the Stokes pulse train. (c) The long-term output powers for the pump and Stokes beams.

We first characterize the stability of the pump and Stokes beams. Their real-time pulse trains, recorded by the oscilloscope at a 1 Gs/s sample rate, are displayed in Figs. 3(a) and 3(b), respectively. Each beam is detected with a 1 GHz photodetector in this measurement before the dichroic mirror. The oscilloscope's online data analysis returns the root-mean-square (RMS) relative pulse-to-pulse peak intensity variations of 0.1% and 0.5%, respectively, for the pump and Stokes (their relative intensity noise spectra are given in Supporting Information). The relatively larger intensity variation of the Stokes beam is mainly from the electrical pulse generator and the intensity modulator. Nevertheless, such an intensity variation is tolerable for CARS measurements, considering the linear dependency of CARS signals on the Stokes light intensity. To evaluate the power stability over a long period, which matters for CARS imaging applications,[6] we continuously monitor the average power for each beam with a power meter for over 60 min. Thanks to the all-fiber configuration, the Stokes beam exhibits excellent power stability (red line in Figure 3c) with an RMS power fluctuation of only 0.2%. A larger power fluctuation (0.5%) is measured for the pump (blue line in Figure 3c), which is mainly due to the mechanical instability of space-to-fiber coupling. Despite that, the overall power stabilities are at the same level as the mode-locked fiber lasers.[31]

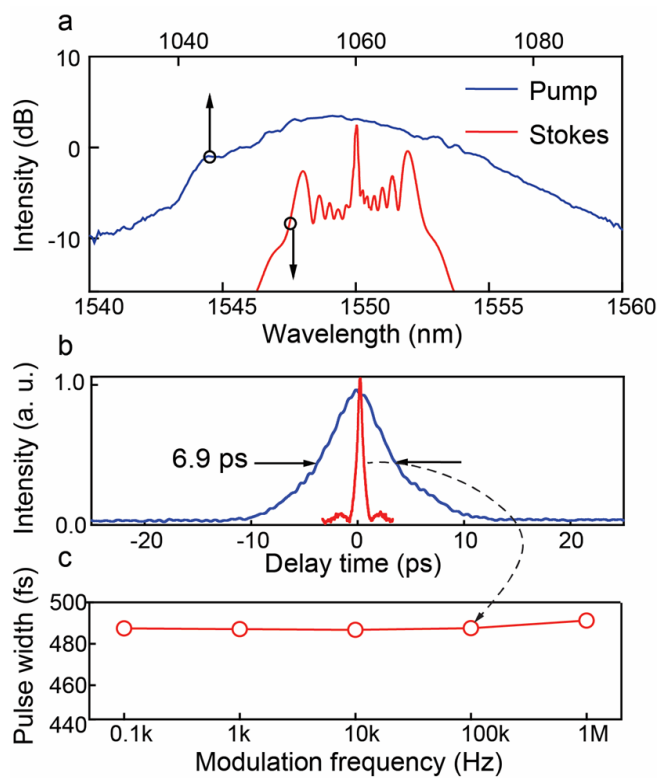

**Figure 4.** The spectral and temporal characterizations of the pump and Stokes beams. (a) The spectra and (b) autocorrelation traces for the pump and Stokes pulses. (c) The pulse width measured at different FM frequencies.

We then characterize the spectral and temporal properties of the two beams. To this end, the beams are directed into an optical spectral analyzer (AQ 6375, Yokogawa) via a silver mirror temporarily placed after the long-pass filter. The recorded spectra are shown in Figure 4a. The modulated structure of the Stokes spectrum (red curve) is due to self-phase modulation in the HNLF. The pump spectrum (blue curve) spans 1035-1080 nm or 9260-9662 $cm^{-1}$ (3-dB bandwidth of 200 $cm^{-1}$), and the Stokes centers at 1550 nm with -3dB spectral width of 5.2 nm (or 21.6 $cm^{-1}$), which together constitute a Raman window of approximately 2800-3200 $cm^{-1}$. This high-wavenumber C-H stretching region is particularly interesting for identifying fats and lipids, the important objects for biomedical studies.[6-8] Figure 4b shows their temporal autocorrelation traces, measured by an optical autocorrelator (pulseCheck, APE). The pump pulses (blue) have a linearly chirped FWHM pulse duration of 6.9 ps, while the Stokes pulses have a 0.49 ps duration, close to transform-limited (assuming a Gaussian shape). The sidelobes for the Stokes beam (red curve) in Figure 4b are due to uncontrolled high-order dispersion. Also, as shown in Figure 4c, the Stokes pulse duration remains unchanged at varied FM frequencies, from 0.1 kHz up to 1 MHz (limited by the RF synthesizer), manifesting the stable performance of the EO pulses during the FM process.

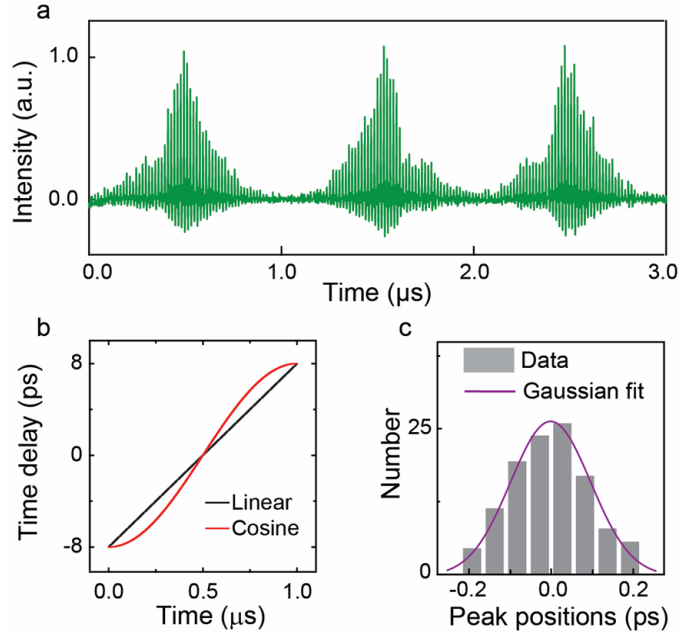

**Figure 5.** Results of time jitter measurements. (a) Cross-correlation traces between the pump and the FM Stokes pulses and (b) the transform functions from the measurement time to the relative time delay. (c) The histogram of the peak positions of the cross-correlation traces.

We also investigate the time jitter between the pump and FM Stokes pulses. For this measurement, we set the EO comb in the sinusoidal FM mode with $f_m$=500 kHz, corresponding to the refresh rate of 1 MHz, and the peak frequency deviation $\Delta f$=7.58 kHz. We then focus the two beams into a 1-mm-thick β-barium borate (BBO) crystal, yielding a sum-frequency cross-correlation signal varying with the measurement time, as shown in Figure 5a. According to Eq. (2), the measurement time ($t$) can be transformed into the relative delay time ($\tau$) via a cosine function, as exemplified by the red curve in Figure 5b. As a comparison, a transformation function for square-wave FM pulses (using the same $f_m$ and $\Delta f$) is plotted in black in Figure 5b. Although square-wave modulation has the advantage of linearity, it is difficult to achieve a high modulation frequency in practice due to the FM bandwidth limitation (square wave generation requires much higher bandwidth than a sine wave). In our case, the maximum modulation frequency allowed by the RF synthesizer is 20 kHz in square-wave FM mode, while it reaches 1 MHz in sinusoidal FM mode, thus suitable for high-speed measurements. For evaluating the time jitter, we statistically analyzed 160 cross-correlation trances against the relative delay time. In Figure 5c, the histogram of the peak position variations of the traces shows a FWHM of 0.22 ps, corresponding to a spectral uncertainty of 6.4 cm$^{-1}$ (estimated using the ratio of the pump spectral width and pulse duration).

*4.2 High-speed CARS measurements*

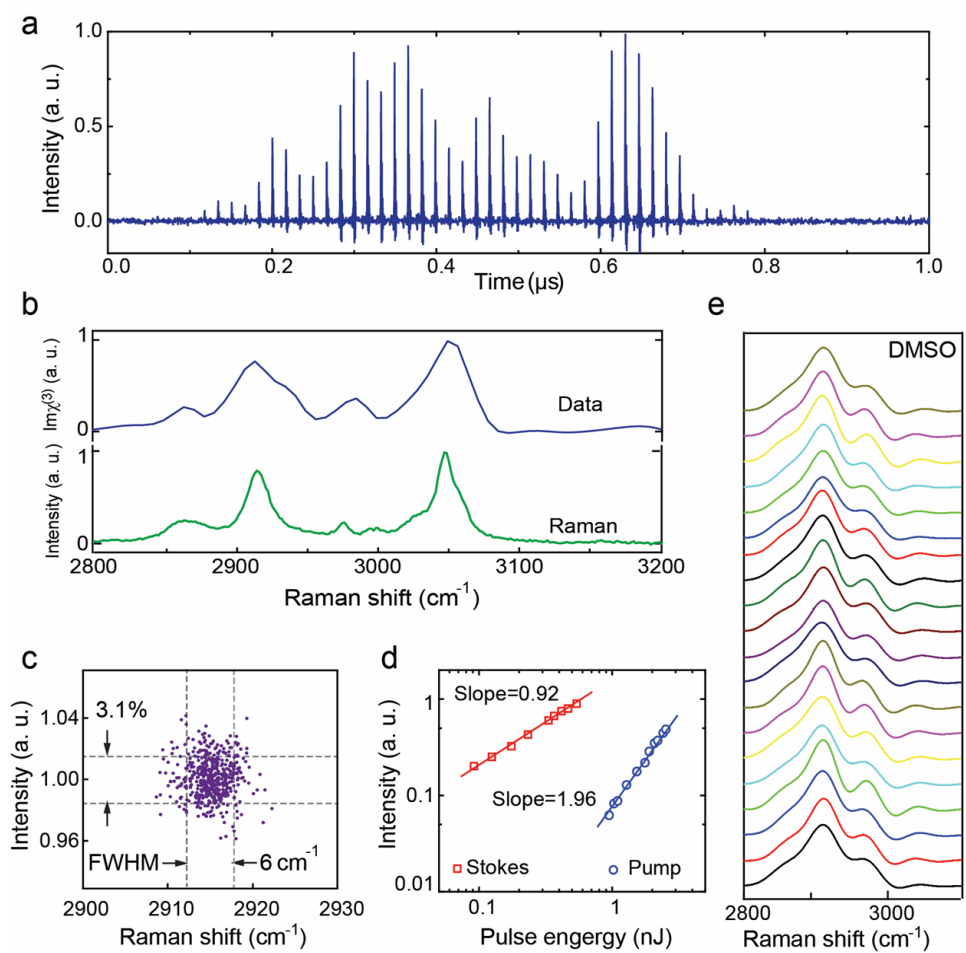

**Figure 6.** Results of CARS measurements. (a) The time-domain CARS sequences (for toluene) and (b) the calibrated spectral data and the corresponding spontaneous Raman spectrum (resolution of 10 cm$^{-1}$) of toluene for comparison. (c) The peak intensity and position variations for the toluene Raman line at 2916 cm$^{-1}$. (d) The CARS signal depending on excitation pulse energies. (e) Consecutive dual-comb measurements for DMSO. For this measurement, the detector signals are electronically low-pass filtered.

We perform multiplex CARS measurement with our hybrid dual-comb source. The raw measurement data, recorded by the oscilloscope (1 Gs/s sample rate), are displayed in Figure 6a. The spectral information is coded in a train of anti-Stokes pulses in the measurement time domain. The sample of interrogation is neat toluene which exhibits distinct Raman characteristics in the high-wavenumber region and therefore is ideal for evaluating the feasibility and fidelity of our method. The dwell time for the spectrum is 1 μs with a duty cycle of ~100% at an acquisition speed of 1 MHz. In Figure 6a, each time-domain pulse (displayed in blue), resembling a narrow line, represents a spectral element for CARS spectroscopy. The SNR for the strongest pulse is 50, calculated by the ratio between the pulse peak and the SD of the noise floor. As expected, the SNR improves with the square root of the data averaging number (Supporting Information).

For retrieving the CARS lines (i.e., the imaginary part of χ$^{(3)}$), we join the pulse maxima in Figure 6a and apply the maximum entropy method[32] using a background spectrum measured for glass (see Supporting Information). We also calibrate the CARS spectrum in wavenumber by multiplying the relative delays by the pump chirp rate and then applying a shift to the wavenumber axis according to the known Raman line at 3056 cm$^{-1}$ (the C-H stretching mode). The retrieved CARS spectrum (blue) is shown in Figure 6b, together with a spontaneous Raman spectrum (green curve) recorded for comparison. Here, the spectral resolution is ~30 cm$^{-1}$ (obtained from the FWHM of the measured CARS lines), which is mainly determined by the Stokes spectral width (21.6 cm$^{-1}$) and the chirp of the pump during the Stokes' pulse duration (i.e., 14.5 cm$^{-1}$). As compared to other systems with spectral resolutions varying from 4 to >100 cm$^{-1}$ (depending on measurement times and Raman spectral widths),[8] this is a moderate resolution, suitable for investigating liquid chemicals and biomedical samples,[6] such as lipids, fats, sterols, etc.

Furthermore, we statistically analyze the Raman peak positions for another line at 2916 cm$^{-1}$ from 200 measurements. The peak positions are plotted in Figure 6c with a FWHM variation of 6 cm$^{-1}$, consistent with the measured time jitter (6.4 cm$^{-1}$). Also, in Figure 6c, the FWHM of the relative intensity variations for the same line of different measurements is 3.1% (or ±1.5%), similar to the other dual-comb systems.[28] Meanwhile, we show the dependences of the CARS signal at 3056 cm-1 on the pump and Stokes pulse energies in Figure 6d. The data are linearly fitted in a log-log scale. The fitting lines indicate that the CARS signal relies linearly on the Stokes beam (with a slope of 0.92, close to 1) and quadratically on the pump beam (with a slope of 1.96, close to 2), in line with our expectations (Supporting Information).

For demonstrating the reproducibility of our method, we plot 20 consecutive raw spectra for pure dimethyl sulfoxide (DMSO) liquid in Figure 6e. For this measurement, we use an electronic low-pass filter (cutoff at 10 MHz) after the photodetector to obtain the spectral envelopes and set the sampling rate of the digitizer to 100 Ms/s to reduce the data size. Figure 7a shows a retrieved CARS spectrum of DMSO (red).

*4.3 Hyperspectral imaging*

We demonstrate hyperspectral imaging using a thin microchip (100-μm thickness) with a Y-shape microchannel (300-μm width). We first fill the microchannel with DMSO and then inject flax seed oil into the channel, forming immiscible oil droplets. The two samples exhibit overlapped spectral features (Figure 7a), highlighting the necessity of hyperspectral measurements. Figure 7b exemplifies a CARS image of the microchip using the line intensities at ~2913 cm$^{-1}$ for DMSO and at ~2850 cm$^{-1}$ for the flax seed oil. In this experiment, the microchip is mounted on a motorized two-dimensional translation stage (Thorlabs, PDX1) for pointwise scanning imaging. The scanning step size is 2 μm for each direction, which defines the spatial resolution. For each point (or pixel), a multiplexed CARS spectrum, similar to these in Figure 7a, is measured with 13 effective elements (spectral span: 400 cm$^{-1}$, resolution: 30 cm$^{-1}$). At each pixel, a multiplexed spectrum measured within 1 μs is recorded. However, limited by the moving stage, the pixel dwell time is ~100 μs (equivalent to 7.7 μs per spectral elements). This case differs from those in which imaging speeds are limited by spectral acquisition processes.[4, 19] In other words, our imaging speed can be improved with fast-moving stages or galvanometric scanners. A technical comparison between our method and the existing high-speed CARS techniques is discussed in Supporting Information, which demonstrates the high-speed advantage of our system in the high-wavenumber region.

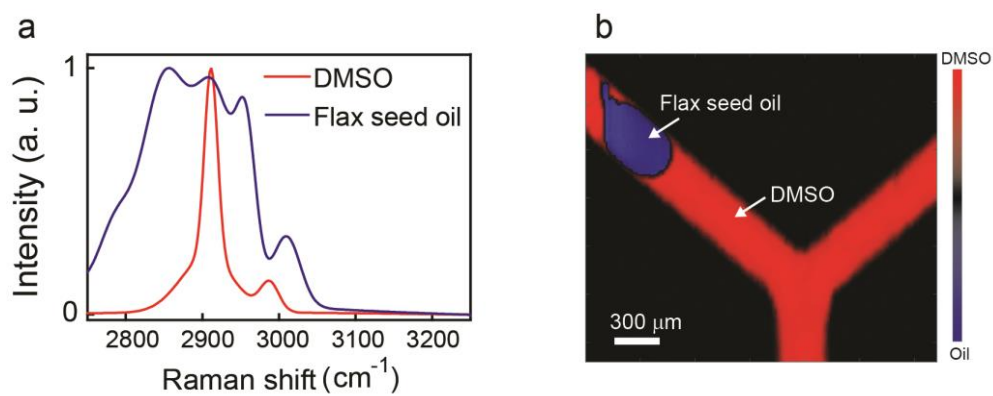

**Figure 7.** Hyperspectral CARS imaging. (a) The normalized dual-comb CARS spectra for DMSO and flax seed oil, respectively, and (b) the spectral image of flax seed oil droplets immerged in the liquid DMSO. The image contains 650 × 650 pixels in total.

## 5. Conclusion

We demonstrate broadband CARS spectroscopy with fast asynchronous spectral tuning, using a linearly chirped fiber comb (spanning 1035-1080 nm) and an FM Stokes EO comb at 1550 nm, with a small relative timing jitter (<0.2 ps) between the combs. We obtain coherent Raman spectra covering 2800-3200 $cm^{-1}$ with a moderate resolution of 30 $cm^{-1}$, and achieve a record acquisition speed of one million spectra per second. This speed (i.e., 1 MHz) can still be increased by using an RF synthesizer with a larger FM bandwidth, but at the cost of reduced spectral elements. Besides, the method can be improved in several aspects. For instance, although our setup is currently limited in the high-wavenumber C-H stretching region, it has the possibility of accessing the important molecular fingerprint region (600-1800 $cm^{-1}$), by further extending the pump light to higher wavelengths (e.g., 1210-1420 nm) via spectral broadening in a nonlinear fiber. Also, it is possible to transfer the excitation wavelengths into the visible or near-infrared regions (e.g., ~775 nm) through optical frequency doubling or sum frequency generation in nonlinear crystals, considering the wavelength-dependent Raman scattering efficiency and the spatial resolutions for imaging. A photomultiplier tube may be used for detecting CARS signals with improved sensitivity. Here, the method is demonstrated for vibrational spectroscopic imaging of neat liquid chemicals and thin films, with various possible applications, from flow cytometry to compositional imaging of materials. More importantly, the concept of using EO synthesis of a continuous-wave laser, in combination with fiber laser technology, for CARS measurements may benefit biomedical applications, such as high-speed label-free imaging, considering the speed advantage, the matureness of RF electronic technology, as well as the flexibility, simplicity, and robustness of the compact EO system. Therefore, our method is of interest for many chemical and biological applications.

**Acknowledgements**

This work is funded in part by the National Nature Science Fund of China (12022411, 61875243, and 62035005), Natural Science Foundation of Chongqing (2022NSCQ-JQX0016), Shanghai Pilot Program for Basic Research (TQ20220104), and Fundamental Research Funds for the Central Universities.

**Author information**

**Corresponding Authors**

**Ming Yan** - *State Key Laboratory of Precision Spectroscopy*, *East China Normal University*, *Shanghai* 200062, *China*; *Chongqing Key Laboratory of Precision Optics*, *Chongqing Institute of East China Normal University*, *Chongqing* 401120, *China*;

Email: myan@lps.ecnu.edu.cn;

**Heping Zeng** - *State Key Laboratory of Precision Spectroscopy*, *East China Normal University*, *Shanghai* 200062, *China*; *Chongqing Key Laboratory of Precision Optics*, *Chongqing Institute of East China Normal University*, *Chongqing* 401120, *China*; *Jinan Institute of Quantum Technology*, *Jinan*, *Shandong* 250101, *China*; *Chongqing Institute for Brain and Intelligence*, *Guangyang Bay Laboratory*, *Chongqing* 400064, *China*;

Email: hpzeng@phy.ecnu.edu.cn;

**Authors**

**Tianjian Lv**- *State Key Laboratory of Precision Spectroscopy*, *East China Normal University*, *Shanghai* 200062, *China*;

**Bing Han** - *State Key Laboratory of Precision Spectroscopy*, *East China Normal University*, *Shanghai* 200062, *China*;

**Zhaoyang Wen** - *State Key Laboratory of Precision Spectroscopy*, *East China Normal University*, *Shanghai* 200062, *China*;

**Kun Huang** - *State Key Laboratory of Precision Spectroscopy*, *East China Normal University*, *Shanghai* 200062, *China*; *Chongqing Key Laboratory of Precision Optics*, *Chongqing Institute of East China Normal University*, *Chongqing* 401120, *China*;

**Kangwen Yang** - *Shanghai Key Lab of Modern Optical System, University of Shanghai for Science and Technology, Shanghai 200093, China*

**Contributions**

M. Y. and H. Z. designed research; T. L., B. H., and Z. W. performed the experiments; T. L., B. H. and M. Y. analyzed data; and T. L. wrote the paper; M. Y., K. H., K. Y. and H. Z. revised the paper.

**Competing financial interests**

The authors declare no competing financial interests.

**Data Availability Statement**

The data that support the findings of this study are available from the corresponding author upon reasonable request.